\documentclass[aps,prl,twocolumn,showpacs,preprintnumbers,amsmath,amssymb,superscriptaddress]{revtex4}

\usepackage{graphicx}
\usepackage{dcolumn}
\usepackage{bm}
\usepackage{lineno}
\usepackage{setspace}
\usepackage{multirow}

\input{defs.tex}

\begin{document}

\title{First Observation of Heavy Baryons \Sib and $\Sib^*$}

\affiliation{Institute of Physics, Academia Sinica, Taipei, Taiwan 11529, Republic of China} 
\affiliation{Argonne National Laboratory, Argonne, Illinois 60439} 
\affiliation{Institut de Fisica d'Altes Energies, Universitat Autonoma de Barcelona, E-08193, Bellaterra (Barcelona), Spain} 
\affiliation{Baylor University, Waco, Texas  76798} 
\affiliation{Istituto Nazionale di Fisica Nucleare, University of Bologna, I-40127 Bologna, Italy} 
\affiliation{Brandeis University, Waltham, Massachusetts 02254} 
\affiliation{University of California, Davis, Davis, California  95616} 
\affiliation{University of California, Los Angeles, Los Angeles, California  90024} 
\affiliation{University of California, San Diego, La Jolla, California  92093} 
\affiliation{University of California, Santa Barbara, Santa Barbara, California 93106} 
\affiliation{Instituto de Fisica de Cantabria, CSIC-University of Cantabria, 39005 Santander, Spain} 
\affiliation{Carnegie Mellon University, Pittsburgh, PA  15213} 
\affiliation{Enrico Fermi Institute, University of Chicago, Chicago, Illinois 60637} 
\affiliation{Comenius University, 842 48 Bratislava, Slovakia; Institute of Experimental Physics, 040 01 Kosice, Slovakia} 
\affiliation{Joint Institute for Nuclear Research, RU-141980 Dubna, Russia} 
\affiliation{Duke University, Durham, North Carolina  27708} 
\affiliation{Fermi National Accelerator Laboratory, Batavia, Illinois 60510} 
\affiliation{University of Florida, Gainesville, Florida  32611} 
\affiliation{Laboratori Nazionali di Frascati, Istituto Nazionale di Fisica Nucleare, I-00044 Frascati, Italy} 
\affiliation{University of Geneva, CH-1211 Geneva 4, Switzerland} 
\affiliation{Glasgow University, Glasgow G12 8QQ, United Kingdom} 
\affiliation{Harvard University, Cambridge, Massachusetts 02138} 
\affiliation{Division of High Energy Physics, Department of Physics, University of Helsinki and Helsinki Institute of Physics, FIN-00014, Helsinki, Finland} 
\affiliation{University of Illinois, Urbana, Illinois 61801} 
\affiliation{The Johns Hopkins University, Baltimore, Maryland 21218} 
\affiliation{Institut f\"{u}r Experimentelle Kernphysik, Universit\"{a}t Karlsruhe, 76128 Karlsruhe, Germany} 
\affiliation{High Energy Accelerator Research Organization (KEK), Tsukuba, Ibaraki 305, Japan} 
\affiliation{Center for High Energy Physics: Kyungpook National University, Taegu 702-701, Korea; Seoul National University, Seoul 151-742, Korea; SungKyunKwan University, Suwon 440-746, Korea} 
\affiliation{Ernest Orlando Lawrence Berkeley National Laboratory, Berkeley, California 94720} 
\affiliation{University of Liverpool, Liverpool L69 7ZE, United Kingdom} 
\affiliation{University College London, London WC1E 6BT, United Kingdom} 
\affiliation{Centro de Investigaciones Energeticas Medioambientales y Tecnologicas, E-28040 Madrid, Spain} 
\affiliation{Massachusetts Institute of Technology, Cambridge, Massachusetts  02139} 
\affiliation{Institute of Particle Physics: McGill University, Montr\'{e}al, Canada H3A~2T8; and University of Toronto, Toronto, Canada M5S~1A7} 
\affiliation{University of Michigan, Ann Arbor, Michigan 48109} 
\affiliation{Michigan State University, East Lansing, Michigan  48824} 
\affiliation{University of New Mexico, Albuquerque, New Mexico 87131} 
\affiliation{Northwestern University, Evanston, Illinois  60208} 
\affiliation{The Ohio State University, Columbus, Ohio  43210} 
\affiliation{Okayama University, Okayama 700-8530, Japan} 
\affiliation{Osaka City University, Osaka 588, Japan} 
\affiliation{University of Oxford, Oxford OX1 3RH, United Kingdom} 
\affiliation{University of Padova, Istituto Nazionale di Fisica Nucleare, Sezione di Padova-Trento, I-35131 Padova, Italy} 
\affiliation{LPNHE, Universite Pierre et Marie Curie/IN2P3-CNRS, UMR7585, Paris, F-75252 France} 
\affiliation{University of Pennsylvania, Philadelphia, Pennsylvania 19104} 
\affiliation{Istituto Nazionale di Fisica Nucleare Pisa, Universities of Pisa, Siena and Scuola Normale Superiore, I-56127 Pisa, Italy} 
\affiliation{University of Pittsburgh, Pittsburgh, Pennsylvania 15260} 
\affiliation{Purdue University, West Lafayette, Indiana 47907} 
\affiliation{University of Rochester, Rochester, New York 14627} 
\affiliation{The Rockefeller University, New York, New York 10021} 
\affiliation{Istituto Nazionale di Fisica Nucleare, Sezione di Roma 1, University of Rome ``La Sapienza," I-00185 Roma, Italy} 
\affiliation{Rutgers University, Piscataway, New Jersey 08855} 
\affiliation{Texas A\&M University, College Station, Texas 77843} 
\affiliation{Istituto Nazionale di Fisica Nucleare, University of Trieste/\ Udine, Italy} 
\affiliation{University of Tsukuba, Tsukuba, Ibaraki 305, Japan} 
\affiliation{Tufts University, Medford, Massachusetts 02155} 
\affiliation{Waseda University, Tokyo 169, Japan} 
\affiliation{Wayne State University, Detroit, Michigan  48201} 
\affiliation{University of Wisconsin, Madison, Wisconsin 53706} 
\affiliation{Yale University, New Haven, Connecticut 06520} 
\author{T.~Aaltonen}
\affiliation{Division of High Energy Physics, Department of Physics, University of Helsinki and Helsinki Institute of Physics, FIN-00014, Helsinki, Finland}
\author{A.~Abulencia}
\affiliation{University of Illinois, Urbana, Illinois 61801}
\author{J.~Adelman}
\affiliation{Enrico Fermi Institute, University of Chicago, Chicago, Illinois 60637}
\author{T.~Affolder}
\affiliation{University of California, Santa Barbara, Santa Barbara, California 93106}
\author{T.~Akimoto}
\affiliation{University of Tsukuba, Tsukuba, Ibaraki 305, Japan}
\author{M.G.~Albrow}
\affiliation{Fermi National Accelerator Laboratory, Batavia, Illinois 60510}
\author{S.~Amerio}
\affiliation{University of Padova, Istituto Nazionale di Fisica Nucleare, Sezione di Padova-Trento, I-35131 Padova, Italy}
\author{D.~Amidei}
\affiliation{University of Michigan, Ann Arbor, Michigan 48109}
\author{A.~Anastassov}
\affiliation{Rutgers University, Piscataway, New Jersey 08855}
\author{K.~Anikeev}
\affiliation{Fermi National Accelerator Laboratory, Batavia, Illinois 60510}
% ...
\collaboration{CDF Collaboration\footnote{With visitors from $^a$University of Athens, 15784 Athens, Greece, 
$^b$University of Bristol, Bristol BS8 1TL, United Kingdom, 
$^c$University Libre de Bruxelles, B-1050 Brussels, Belgium, 
$^d$Cornell University, Ithaca, NY  14853, 
$^e$University of Cyprus, Nicosia CY-1678, Cyprus, 
$^f$University College Dublin, Dublin 4, Ireland, 
$^g$University of Edinburgh, Edinburgh EH9 3JZ, United Kingdom, 
$^h$University of Heidelberg, D-69120 Heidelberg, Germany, 
$^i$Universidad Iberoamericana, Mexico D.F., Mexico, 
$^j$University of Manchester, Manchester M13 9PL, England, 
$^k$Nagasaki Institute of Applied Science, Nagasaki, Japan, 
$^l$University de Oviedo, E-33007 Oviedo, Spain, 
$^m$University of London, Queen Mary College, London, E1 4NS, England, 
$^n$University of California Santa Cruz, Santa Cruz, CA  95064, 
$^o$Texas Tech University, Lubbock, TX  79409, 
$^p$University of California, Irvine, Irvine, CA  92697, 
$^q$IFIC(CSIC-Universitat de Valencia), 46071 Valencia, Spain, 
}}
\noaffiliation

\begin{abstract}
We report an observation of new bottom baryons produced 
in $p\bar{p}$ collisions at the Tevatron.  Using $1.1~\rm{ fb}^{-1}$  
of data collected by the  \cdf2 detector, we observe four 
$\Lb\pi^{\pm}$ resonances in the fully reconstructed decay mode 
$\Lb \ra \Lc \pi^-$, where $\Lc \ra p K^- \pi^+$.  
We interpret these states as the $\Sigma_b^{(*)\pm}$ baryons and 
measure the following masses:
\begin{center}
$m_{\Sib^+} = 5807.8^{+2.0}_{-2.2}~{\rm (stat.)} \pm 1.7~{\rm (syst.)}~\mevcc$ \\
$m_{\Sib^-} = 5815.2 \pm 1.0~{\rm (stat.)} \pm 1.7~{\rm (syst.)}~\mevcc$ \\
$m(\Sib^*) - m(\Sib) = 21.2^{+2.0}_{-1.9}~{\rm (stat.)} ^{+0.4}_{-0.3}~{\rm (syst.)}~\mevcc$\\
\end{center}
\end{abstract}

 \pacs{14.20.Mr, 13.30.Eg}
 \keywords{B physics, beautiful baryons}

\maketitle

Recently the \cdf2 detector at the Fermilab Tevatron
has accumulated the world's largest sample of fully reconstructed 
$\Lb$ baryons, which consist of the $u$, $d$, and $b$ quarks,
with $3180 \pm 60$ (stat.) $\Lb\ra\Lcpi$ candidates.
This is made possible by the large
$b\bar{b}$ production cross-section
in $p\bar{p}$ collisions at $\sqrt{s} = 1.96$~TeV
and the ability of the \cdf2 experiment to select fully 
hadronic decays of $b$ hadrons with a specialized trigger system.
In this Letter, we present an observation of four 
$\Lb\pi^{\pm}$ resonances,
where $\Lb\ra\Lc\pi^-$ and $\Lc \ra p K^- \pi^+$,
using 1.1 fb$^{-1}$ of data.
The $\Lb\pi$ states are interpreted as the lowest-lying 
charged \Sib baryons and will be labeled $\Sib^{(*)}$.
The symbol $\Sib$ refers to $\Sib^{\pm}$, while $\Sib^*$ refers to $\Sib^{*\pm}$.
Any reference to a specif\mbox{}ic 
charge state implies the
antiparticle state as well.

The $\Sib^{(*)+}$ baryons contain one $b$ and two $u$ quarks, 
while the $\Sib^{(*)-}$ baryons contain one $b$ and two $d$ quarks;
these states are expected to exist but have not been observed.
Baryons containing one bottom quark and two light quarks 
can be described by heavy quark ef\mbox{}fective
theory (HQET)~\cite{Manohar:2000dt}.
In HQET a bottom baryon consists of a $b$
quark acting as a static source of the color f\mbox{}ield  surrounded by 
a diquark system comprised of the two light quarks.
In the lowest-lying $\Sib^{(*)}$ states, 
the diquark system has strong isospin $I=1$
and $J^P=1^+$, which couple to the heavy quark spin and result in a
doublet   of    baryons   with   $J^P=\frac{1}{2}^+$    ($\Sib$)   and
$J^P=\frac{3}{2}^+$   ($\Sib^*$). 
This doublet is degenerate for inf\mbox{}inite $b$ quark mass. 
As the $b$ quark mass is f\mbox{}inite, there is a
hyperf\mbox{}ine mass splitting between the $\frac{3}{2}^+$ and $\frac{1}{2}^+$ 
states.  There is  also an
isospin mass splitting between the $\Sib^{(*)-}$ and $\Sib^{(*)+}$ states.

Predictions for the $\Sib^{(*)}$ masses come
from non-relativistic and relativistic potential quark models 
\cite{Stanley:1980fe},
$1/N_c$ expansion~\cite{Jenkins:1996de},
quark models in the HQET approximation 
\cite{Albertus:2003sx}, sum rules~\cite{Roncaglia:1995az},
and lattice quantum chromodynamics calculations
\cite{Bowler:1996ws}.  On the basis of 
\cite{Stanley:1980fe,Jenkins:1996de,Albertus:2003sx,Roncaglia:1995az,
Bowler:1996ws},
we expect $m(\Sib) - m(\Lb) \sim 180 - 210~\mevcc$, 
$m(\Sib^*) - m(\Sib) \sim 10 - 40~\mevcc$, and 
$m(\Sib^{(*)-}) - m(\Sib^{(*)+}) \sim 5 - 7~\mevcc$.
The dif\mbox{}ference between the isospin mass splittings of the $\Sib^{*}$ 
and $\Sib$ multiplets is predicted to be 
$[m(\Sib^{*+})-m(\Sib^{*-})] - [m(\Sib^+)-m(\Sib^-)] = 0.40 \pm 0.07~\mevcc$
\cite{Rosner:1998zc}.
The natural width of $\Sib^{(*)}$ baryons is expected to be dominated 
by the P-wave one pion transition $\Sigma_b^{(*)}\ra\Lb\pi$, whose 
partial width depends on the available phase space and 
the pion coupling to a constituent quark.
Using an HQET prediction~\cite{Korner:1994nh}, the natural
widths for the expected $\Sib^{(*)}$ masses are
$\Gamma(\Sib^{\pm}) \approx 7~\mevcc$ and
$\Gamma(\Sib^{*\pm}) \approx 13~\mevcc$.


The \cdf2 detector is described in detail elsewhere~\cite{Acosta:2004yw}. 
Its components and capabilities most relevant to this analysis are the 
tracking system~\cite{Sill:2000zz}
and a displaced track trigger which is employed to select bottom and 
charmed hadrons~\cite{ttt}.

In reconstructing the decays $\Lb \ra \Lc \pi^-$ and $\Lc \ra p K^- \pi^+$,
the proton from the \Lc decay and the $\pi^-$ from the \Lb decay
both must have \pt $>$ 2 $\gevc$~\cite{det_coord},
while the $K^-$ and $\pi^+$ candidates have $\pt > 0.5~\gevc$.
We also require $\pt(p)>~\pt(\pi^{+})$ to suppress \Lc 
combinatorial background.
No particle identification is used in this analysis.  All
particle hypotheses consistent with the candidate decay
structure are considered.
In a 3-D kinematic f\mbox{}it, the \Lc daughter tracks are constrained
to originate from a single point. 
The \Lc candidate is constrained to the known \Lc mass, 
and the \Lc momentum vector is extrapolated to intersect 
the $\pi^-$ momentum vector to form the \Lb vertex.  
The probability of the 3-D \Lb kinematic vertex f\mbox{}it must exceed 
$0.1\%$, and the \Lc and \Lb must have \pt greater than $4.5$ 
and $6.0~\gevc$, respectively.
To suppress prompt backgrounds from the primary interaction, we make
the following decay time requirements:
$\ct(\Lb) > 250~\mkm$ and its signif\mbox{}icance $\ct(\Lb)/\sigma_{\ct}>10$.
We define $\ct(\Lb) \equiv  L_{\rm xy}(\Lb) m_{\Lb}c/\pt(\Lb)$ as
the \Lb proper time, 
where $L_{\rm xy}(\Lb)$ is def\mbox{}ined as the 
projection, onto $\pt(\Lb)$, of the vector
connecting the primary vertex to the \Lb vertex in the transverse plane.  
We use a primary vertex determined event-by-event 
when computing this vertex displacement.
To reduce combinatorial backgrounds and partially reconstructed decays, 
we also require $\left |d_{0}(\Lb)\right | < 80~\mkm$,
where $d_0(\Lb)$ is the impact parameter of the momentum vector
of the \Lb candidate with respect to the primary vertex.
To suppress the contributions from
$\bar{B}^0 \to D^+\pi^-$ decays, where $D^+ \to K^- \pi^+ \pi^+$,
we require $m(\pkpi)$ to be within 16 $\mevcc$ of the known \Lc 
mass~\cite{pdg}, and $\ct(\Lc) \in [-70, 200]~\mkm$.
We define $\ct(\Lc) \equiv L_{\rm xy}(\Lc) m_{\Lc}c/\pt(\Lc)$
as the \Lc proper time, where $L_{\rm xy}(\Lc)$ is def\mbox{}ined 
analogously to $L_{\rm xy}(\Lb)$ but computed with respect
to the \Lb vertex.

\begin{figure}
\includegraphics[width=0.4\textwidth]{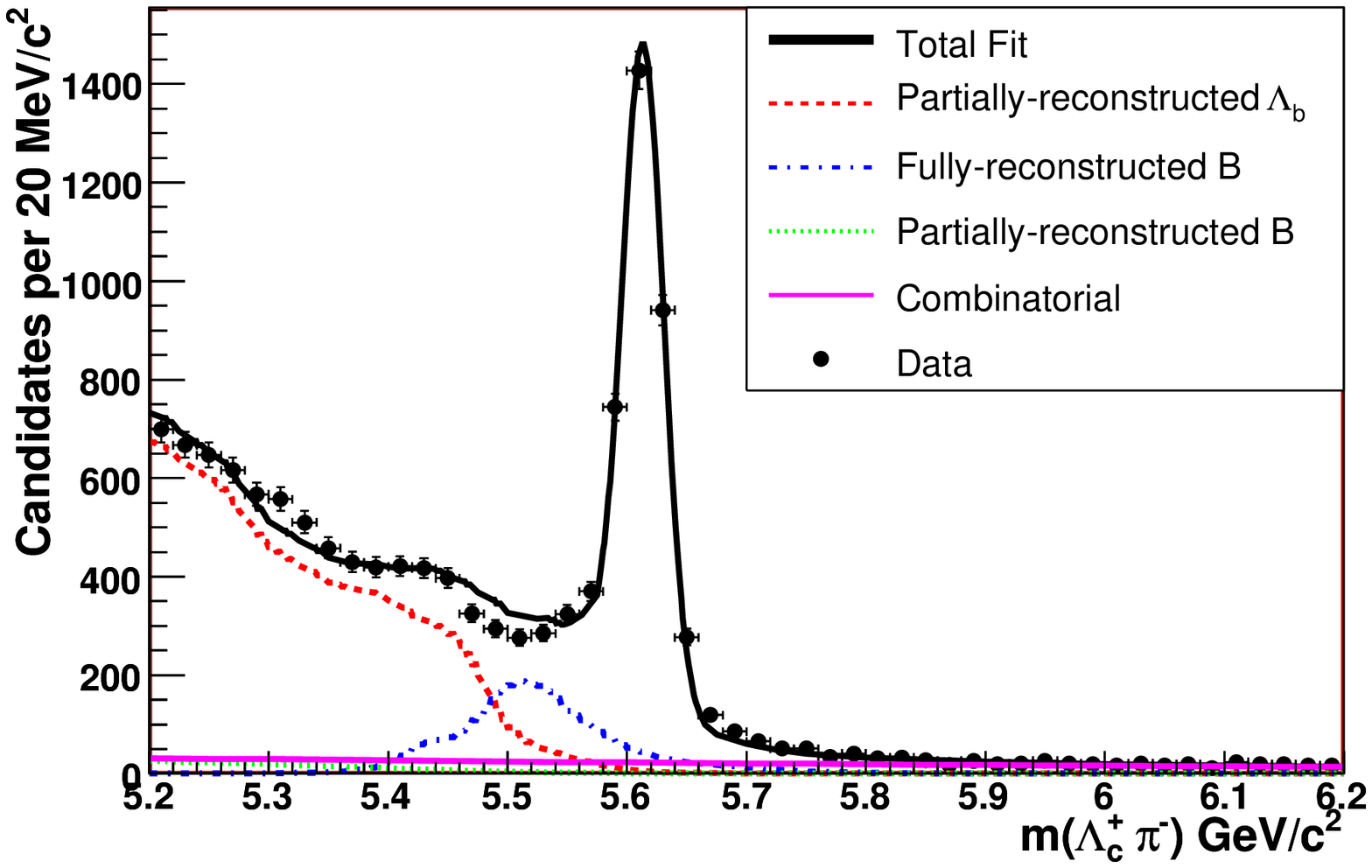}
\caption{Fit to the invariant mass of $\Lb\ra\Lc\pi^-$ candidates.
	Curves for fully reconstructed \Lb decays such as $\Lb\ra\Lcpi$ 
	and $\Lb\ra\Lc K^-$ are not indicated on the f\mbox{}igure.
	The \Lb signal region, $m(\Lcpi) \in [5.565, 5.670]~\gevcc$,
	consists primarily of \Lb baryons, with some contamination
	from $B$ mesons and combinatorial events.
	The discrepancies between the fit and data below the \Lb 
	signal region are due to incomplete knowledge of the branching 
	ratios of the decays in this region and are included
	in the $\Sib^{(*)}$ background model systematics.}
\label{fig:lb_mass}
\end{figure}

The invariant mass distribution of \Lcpi candidates is shown
in Figure \ref{fig:lb_mass}
overlaid with a binned maximum likelihood f\mbox{}it.
A clear $\Lb\to\Lcpi$ signal is observed at the expected \Lb mass.
The invariant mass distribution is described by several
components:  the $\Lb\to\Lcpi$ signal, a combinatorial background,
partially
and fully reconstructed $B$ mesons which pass the \Lcpi selection 
criteria, partially reconstructed \Lb decays, and fully reconstructed 
\Lb decays other than $\Lcpi$ ({\it e.g.} $\Lb \to \Lc K^-$).
The combinatorial background is modeled with an exponentially
decreasing function.
All other components are represented
in the f\mbox{}it by f\mbox{}ixed shapes derived from 
Monte Carlo (MC) simulations~\cite{inclusive-b-samples, Lange:2001uf}.
The normalizations
are constrained by Gaussian terms to branching ratios that are
either measured (for $B$ meson decays) or theoretical predictions 
(for \Lb decays).
In the f\mbox{}it, the \Lb components are normalized
relative to the $\Lb\to\Lcpi$ signal.  To normalize the $B$ meson components, 
we explicitly reconstruct a $\bar{B}^0 \to (K^-\pi^+\pi^+)\pi^-$ signal 
in the \Lcpi sample by replacing the proton mass hypothesis 
with the pion mass hypothesis. 
The f\mbox{}it to the invariant \Lcpi mass distribution
results in $3180 \pm 60$ (stat.) $\Lb\ra\Lcpi$ candidates.


%

The reconstruction of $\Sib^{(*)}$ proceeds by combining \Lb candidates 
in the \Lb signal region, $m(\Lcpi) \in [5.565, 5.670]~\gevcc$, with 
all remaining high quality tracks.  A pion mass hypothesis is used
when computing the invariant mass of the $\Sib^{(*)}$ candidate.
We search for narrow resonances in the mass dif\mbox{}ference distribution 
of $Q = m(\Lb \pi) - m(\Lb) - m_{\pi}$, where $m(\Lb)$ is the 
reconstructed $\Lcpi$ mass.
The $\Sib^{(*)}$ candidates are divided into two subsamples using
the charge of the pion from $\Sib^{(*)}$ decay, denoted by $\piSb$:
in the $\Lb\pi^-$ subsample the $\piSb$ has the same charge as the pion 
from $\Lb$ while in the $\Lb\pi^+$ subsample the $\piSb$ has the opposite 
charge as the pion from $\Lb$. 

The $\Sib^{(*)}$ signal region, def\mbox{}ined as $Q\in[30,100]~\mevcc$,
is motivated by the predictions in 
\cite{Stanley:1980fe,Jenkins:1996de,Albertus:2003sx,Roncaglia:1995az,
Bowler:1996ws}.
The signal is modeled by the {\sc pythia}~\cite{Sjostrand:2000wi} 
event generator where only the decays $\Sib^{(*)}\ra\Lb\pi$, 
$\Lb\ra\Lc\pi^-$, and $\Lc\ra pK^-\pi^+$ are allowed.  
We optimize the $\Sib^{(*)}$ selection criteria by maximizing
$\epsilon(S_{\rm MC})/\sqrt{B}$, where $\epsilon(S_{\rm MC})$ 
is the ef\mbox{}f\mbox{}iciency of the $\Sib^{(*)}$ signal measured 
in the MC simulation and $B$ is the number of background events
in the signal region estimated from the upper and lower sideband   
regions of $Q\in[0, 30]~\mevcc$ and $Q\in[100, 500]~\mevcc$.
These sideband regions are parameterized by 
a power law multiplied by an exponential.  
We combine the \rs~and \ws~subsamples to optimize cuts on the
\pt of the $\Sib^{(*)}$ candidate, the impact parameter signif\mbox{}icance 
\dsig of the \piSb track, and the \costhst of the \piSb track, where
$\theta^*$ is def\mbox{}ined as the angle between the momentum of the 
\piSb in the $\Sib^{(*)}$ rest frame and the direction of the total $\Sib^{(*)}$
momentum in the lab frame.  
The maximum of $\epsilon(S_{\rm MC})/\sqrt{B}$ is realized for
$\ptSb > 9.5~\gevc$, $\dsig < 3.0$,
and $\costhst >-0.35$.  

\begin{figure}
\includegraphics[width=0.40\textwidth]{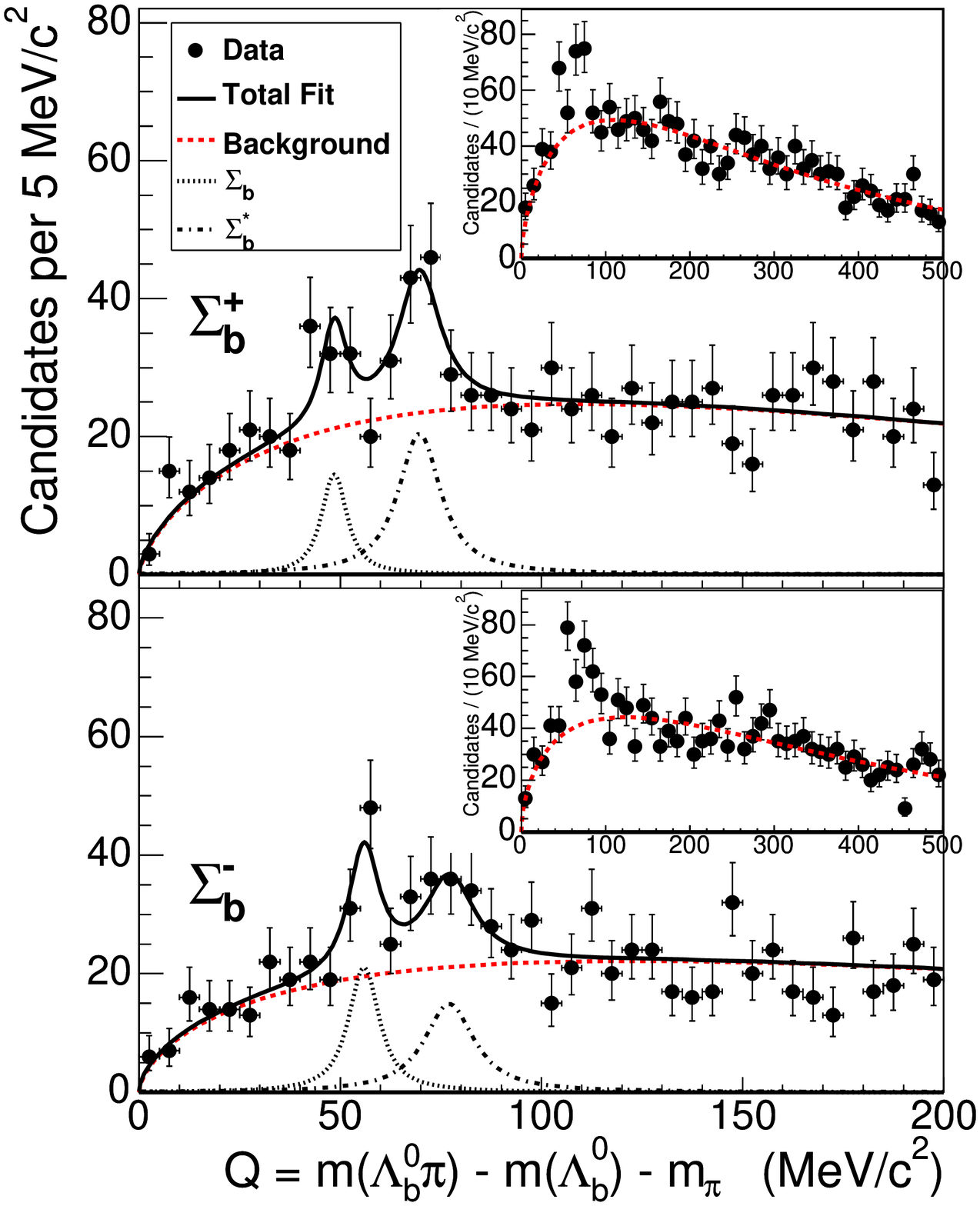}
\caption{The $\Sib^{(*)}$ f\mbox{}it to the \ws~and \rs~subsamples.  The top plot shows
	the \ws~subsample, which contains $\Sib^{(*)+}$, while the
	bottom plot shows the \rs~subsample, which contains $\Sib^{(*)-}$.
  	The insets show the expected background plotted on the data for
	$Q \in$ [0, 500] $\mevcc$, while the signal
	f\mbox{}it is shown on a reduced range of $Q \in$ [0, 200] $\mevcc$.}
\label{fig:combined_Q_plot}
\end{figure}

In the $\Sib^{(*)}$ search, the dominant background is from
the combination of prompt \Lb baryons with extra tracks
produced in the hadronization of the $b$ quark.  The remaining 
backgrounds are from the combination of hadronization tracks
with $B$ mesons reconstructed as \Lb baryons, and from
combinatorial background events. 
The percentage of each background component in the \Lb signal region, 
computed from the \Lb mass f\mbox{}it, 
is $(89.5\pm1.7)$\% $\Lb$ baryons, $(7.2\pm0.6)$\% $B$ mesons, 
and $(3.3\pm0.1)$\% combinatorial events.
Other backgrounds such as 5-track decays of $B^+$ mesons
are negligible, as conf\mbox{}irmed in
inclusive single $b$ hadron simulations~\cite{inclusive-b-samples, Lange:2001uf}.
The high mass region above the $\Lb\ra\Lcpi$ signal in 
Figure \ref{fig:lb_mass} 
determines the combinatorial background shape.
Reconstructing $\bar{B}^0\ra D^+\pi^-$ data as $\Lb\ra\Lc\pi^-$
gives the $B$ \HaUE\ background shape.
The \Lb \HaUE~background shape is obtained from a $\Lb\ra\Lcpi$ 
{\sc pythia} simulation.
The events in this simulation are reweighted so that 
the $\pt(\Lb)$ distribution agrees with data.
As the simulation has fewer low momentum tracks around 
the \Lb than found in data, the simulated events are further 
reweighted until the \pt spectrum of tracks around the \Lb is
consistent with data.  
The background shapes are parameterized by a power law 
multiplied by an exponential, and the normalizations are
fixed from the percentage of that background 
component in the \Lb signal region.
The total background shown in Figure \ref{fig:combined_Q_plot} (inset)
is compatible with the $Q$ sidebands and the background shape
and normalization are f\mbox{}ixed components of the $\Sib^{(*)}$ 
f\mbox{}it.


In the $Q$ signal region we observe an excess of events over the 
total background as shown in Figure \ref{fig:combined_Q_plot}.  
The excess in the \rs\ subsample is $118$
over $288$ expected background candidates.  In the \ws\ subsample 
the excess is $91$ over $313$ expected background candidates.

We perform a simultaneous unbinned maximum likelihood f\mbox{}it to 
the \rs\ and \ws\ subsamples for a signal from each expected $\Sib^{(*)}$ 
state plus the background, referred to as the ``four signal hypothesis.'' 
Each signal consists of a non-relativistic Breit-Wigner distribution
convoluted with two Gaussian distributions describing the detector 
resolution, with a dominant narrow core of an $1.2~\mevcc$ width
and a small broad component of a $3~\mevcc$ width
for the tails.  The natural width of each 
Breit-Wigner distribution is computed from the central $Q$ value
\cite{Korner:1994nh}.
The expected dif\mbox{}ference of the isospin mass splittings 
within the $\Sib^{*}$ and $\Sib$ multiplets
is below our 
sensitivity with this sample of data.  Consequently, we constrain 
$m(\Sib^{*+})-m(\Sib^+) = m(\Sib^{*-})-m(\Sib^-) \equiv \Delta_{\Sib^*}$.
The four \Sib signal f\mbox{}it to data, which has a f\mbox{}it probability
of 76\% in the range $Q\in[0, 200]~\mevcc$, is shown in Figure \ref{fig:combined_Q_plot}.

Systematic uncertainties on the mass dif\mbox{}ference and yield 
measurements fall into three categories: mass scale, 
$\Sib^{(*)}$ background model, and $\Sib^{(*)}$ signal parameterization.
The systematic uncertainty on the mass scale is determined by
the discrepancies of the \cdf2 measured $Q$ values of the $D^*$, $\Sic$,
and $\Lambda_c^*$ hadrons from the world average $Q$ values~\cite{pdg}.
The $Q$ value dependence of this systematic uncertainty is modeled
with a linear function, which is used to extrapolate
the mass scale uncertainty for each $\Sib^{(*)}$ $Q$ value.
This is the largest systematic uncertainty for the mass difference
measurements, ranging from 0.1 to 0.3 $\mevcc$.
The systematic ef\mbox{}fects related to assumptions made on the 
$\Sib^{(*)}$ background model are:
the sample composition of the \Lb signal region, 
the normalization and functional form of the \Lb \HaUE\ background
taken from a {\sc pythia} simulation,
and our limited knowledge of the shape of the \Lb \HaUE\ background
(the largest systematic uncertainty on the yield measurements, ranging
from 2 to 15 events).
The systematic ef\mbox{}fects related to assumptions made on the 
$\Sib^{(*)}$ signal parameterization are:
underestimation of the detector resolution,
the uncertainty in the natural width prediction from~\cite{Korner:1994nh}, 
and the constraint that 
$m(\Sib^{*+})-m(\Sib^+) = m(\Sib^{*-})-m(\Sib^-)$.
%


The signif\mbox{}icance of the signal is evaluated using the
likelihood ratio, $LR \equiv L/ L_{\rm alt}$, where $L$ is the 
likelihood of the four signal hypothesis and $L_{\rm alt}$ is the
likelihood of an alternative hypothesis~\cite{Royall}.
We study the alternate hypotheses of no signal,  two 
\Sib states (one per $\Lb\pi$ charge combination), and 
three $\Sib^{(*)}$ states, performed by eliminating one of
the states in the four signal hypothesis.
Systematic variations are included in the f\mbox{}it as 
nuisance parameters over which the likelihood is integrated.
The resulting likelihood ratios are given in Table \ref{tab:nll_ratios}.
To assess the significance of the signal, we repeat the four signal 
hypothesis f\mbox{}it on samples randomly generated from alternate signal
hypotheses.  In 12 million background samples, none had a $LR$ equivalent 
or greater than the one found in data.  We evaluate
the probability for background only to produce four signals of this or
greater significance to be less than $8.3\times 10^{-8}$, 
corresponding to a signif\mbox{}icance of greater than $5.2~\sigma$.  The
probabilities for each of the alternate hypotheses to produce the observed
signal structure is also given in Table \ref{tab:nll_ratios}.
The f\mbox{}inal results for the \Sib measurement
are quoted in Table \ref{tab:final_values}.
Using the \cdf2 measurement of 
$m_{\Lb} = 5619.7 \pm 1.2~{\rm (stat.)} \pm 1.2~{\rm (syst.)}~\mevcc$
\cite{Acosta:2005mq}, we f\mbox{}ind the absolute masses of the
\Sib states given in Table \ref{tab:final_values}.  The systematic 
uncertainties on the absolute \Sib mass values are dominated by the 
total \Lb mass uncertainty.

\begin{table}[bth]\centering
\caption{Likelihood ratios ($LR$) in favor of the four signal
	hypothesis over alternative hypotheses.  Also shown is 
	the probability for each hypothesis
	to produce the observed data ($p$-value), calculated using
	the $LR$ as a test statistic on randomly generated samples.
	The final column gives the equivalent standard deviations from
	the normal distribution.
	\label{tab:nll_ratios}}
\begin{tabular}{l c c c} 
\hline\hline
Hypothesis		& $LR$ 			& $p$-value		& Significance ($\sigma$)	\\
\hline
No Signal		& $2.6\times 10^{18}$	& $<8.3\times 10^{-8}$	& $>5.2$	\\
Two \Sib States		& $4.4\times 10^{6}$	& $9.2\times10^{-5}$	& $3.7$	\\
No $\Sib^-$ Signal	& $1.2\times 10^{5}$	& $3.2\times 10^{-4}$	& $3.4$	\\
No $\Sib^+$ Signal	& $49$			& $9.0\times 10^{-3}$	& $2.4$	\\
No $\Sib^{*-}$ Signal	& $4.9\times 10^{4}$	& $6.4\times 10^{-4}$	& $3.2$	\\
No $\Sib^{*+}$ Signal	& $8.1\times 10^{4}$	& $6.0\times 10^{-4}$	& $3.2$	\\
\hline\hline
\end{tabular}
\end{table}

\begin{table}[bth]\centering
\caption{Final results for the \Sib measurement.  The first uncertainty
	is statistical and the second is systematic.  The absolute \Sib mass
	values are calculated using a \cdf2 measurement of the \Lb 
	mass~\cite{Acosta:2005mq}, which contributes to the 
	systematic uncertainty.
	\label{tab:final_values}}
\begin{tabular}{l l l l} 
\hline\hline
State	        & \multicolumn{1}{c}{Yield} 
& \multicolumn{1}{c}{$Q$ or $\Delta_{\Sib^*}$ ($\mevcc$)} & \multicolumn{1}{c}{Mass ($\mevcc$)}	\\
\hline
${\Sib^+}$	& $32^{+13+5}_{-12-3}$	& $Q_{\Sib^+} = 48.5^{+2.0+0.2}_{-2.2-0.3}$ 	& $5807.8^{+2.0}_{-2.2}\pm 1.7$	\\
${\Sib^-}$	& $59^{+15+9}_{-14-4}$	& $Q_{\Sib^-} = 55.9\pm 1.0\pm 0.2$		& $5815.2\pm 1.0\pm 1.7$      \\
${\Sib^{*+}}$	& $77^{+17+10}_{-16-6}$ & \multirow{2}{*}{$\Delta_{\Sib^*} = 21.2^{+2.0+0.4}_{-1.9-0.3}$} & $5829.0^{+1.6+1.7}_{-1.8-1.8}$ \\
${\Sib^{*-}}$	& $69^{+18+16}_{-17-5}$ & 	& $5836.4\pm 2.0^{+1.8}_{-1.7}$	\\
\hline\hline
\end{tabular}
\end{table}
%


In summary, using a sample of $3180 \pm 60$ (stat.) $\Lb \to \Lcpi$ candidates
reconstructed in  $1.1~\rm{ fb}^{-1}$ of \cdf2 data, we search for
resonant $\Lb\pi^\pm$ states.  We observe a
signal of four states whose masses and widths are consistent with those
expected for the lowest-lying charged $\Sib^{(*)}$ baryons.
This result represents the f\mbox{}irst observation of the $\Sib^{(*)}$ baryons.


We thank T.~Becher, A.~Falk, D.~Pirjol, J.~Rosner, and D.~Ebert 
for useful discussions.
We also thank the Fermilab staf\mbox{}f and the technical staf\mbox{}fs of the 
participating institutions for their vital contributions. This work 
was supported by the U.S. Department of Energy and National Science 
Foundation; the Italian Istituto Nazionale di Fisica Nucleare; the 
Ministry of Education, Culture, Sports, Science and Technology of Japan; 
the Natural Sciences and Engineering Research Council of Canada; the 
National Science Council of the Republic of China; the Swiss National 
Science Foundation; the A.P. Sloan Foundation; the Bundesministerium f\"ur 
Bildung und Forschung, Germany; the Korean Science and Engineering Foundation 
and the Korean Research Foundation; the Particle Physics and Astronomy 
Research Council and the Royal Society, UK; the Institut National de 
Physique Nucleaire et Physique des Particules/CNRS; the Russian Foundation 
for Basic Research; the Comisi\'on Interministerial de Ciencia y 
Tecnolog\'{\i}a, Spain; the European Community's Human Potential Programme 
under contract HPRN-CT-2002-00292; and the Academy of Finland.


\end{document}